\documentstyle[aps,prl]{revtex}
\input epsf.sty
\begin{document}

\twocolumn[\hsize\textwidth\columnwidth\hsize\csname@twocolumnfalse\endcsname

\draft

\title{Resistivity in the spin-gap state of the t-J model}
\author{Masaru Onoda\cite{onoda}}
\address{Department of Physics, University of Tokyo, 
Hongo, Tokyo, 113 Japan 
}
\author{Ikuo Ichinose\cite{ichinose}}
\address{Institute of Physics, University of Tokyo, 
Komaba, Tokyo, 153 Japan}
\author{Tetsuo Matsui\cite{matsui}}
\address{Department of Physics, Kinki University, 
Higashi-Osaka, 577 Japan
}

\date{\today}

\maketitle

\begin{abstract}

Being motivated by recent experimental data on YBaCuO, we 
calculate dc resistivity $\rho$ in the spin-gap state  of 
charge-spin-separated t-J model by  using a massive gauge theory
of holons and spinons. The result shows $\rho(T)$ deviates 
downward from the $T$-linear behavior below the spin-gap on-set 
temperature $T^*$ as $\rho(T) \propto T\{1 -c(T^*-T)^d\}$ 
where the mean field value of $d$ is 1/2.
To achieve smooth deviation from 
the $T$-linear behavior, one needs $d > 1$.  
The deviation becomes reduced with increasing hole doping.
\end{abstract}


]


Many quasi-two-dimensional cuprates with high superconducting 
transition temperature $T_c$ exhibit anomalous metallic 
behavior above $T_c$ 
in Hall coefficient, magnetic susceptibility, etc. \cite{sato}, which
calls for a new theoretical explanation, probably in  a framework
beyond the conventional Fermi-liquid theory.
Anderson \cite{anderson} pointed out  that  
charge-spin separation (CSS) phenomenon may be a possible explanation.

In the t-J model of strongly-correlated electrons, 
the CSS is naturally described by the slave-boson (SB) 
(or slave-fermion) mean field theory (MFT).
When one incorporates fluctuations of MF's,  they behave as  
gauge fields  
coupled to holons and spinons and the system possesses a U(1) gauge 
symmetry.
The CSS can be interpreted as  a deconfinement phenomenon of 
this gauge theory. The system is expected to have a 
confinement-deconfinement (CD) phase transition  and the transition 
temperature  $T_{\rm CSS}$ can be estimated \cite{imcss},
below which the CSS takes place.

The observed $T$-linear behavior of dc resistivity $\rho(T)$ \cite{fiory}
has been taken as one of  signals of the universal anomaly
in the  metallic phase of high-$T_c$ cuprates. 
Lee and Nagaosa \cite{ln} showed $\rho(T) \propto T$
for fermions and bosons interacting with  a massless gauge field. 
This  system has some relation to the  uniform RVB MFT
of the t-J model in the SB representation.

Recent experiments on YBaCuO by Ito et al.\cite{uchida}
 and others reported that
$\rho(T)$ deviates downward from the $T$-linear behavior below
certain temperature $T^* ( > T_c)$. This  $T^*$  coincides with the  
temperature determined by NMR and neutron experiments \cite{sgexp}
at which a  spin gap starts to develop. 
So it is quite interesting to calculate
$\rho$ in the spin-gap state of  the SB t-J model.

The effective theory used in Ref.\cite{ln} is inadequate
for this purpose, since it assumes no spin gap, containing
only one gauge field associated with the hopping amplitudes of
holons and spinons. 
In Ref.\cite{imcss}, we introduced multiple gauge fields, and
argued that, when a spin gap develops, these gauge fields become massive
due to a gauge version of Anderson-Higgs mechanism.
Below we use  an effective gauge field theory of holons and spinons
emerged from these considerations and calculate  $\rho$ for 
the region $T_c < T < T^*$ to obtain  $\rho \propto T\{1-c(T^*-T)^d\} $, 
near $T^*$ where $d$
is the critical exponent of gauge-boson mass,
$m_A \propto (T^* - T)^d$.
This reduction reflects the fact
that $m_A^2$ suppresses  gauge-field fluctuations and inelastic scatterings 
between gauge bosons and holons and spinons.
The MFT gives $d = 1/2$, but
to obtain a more reliable $\rho$, one needs to calculate
$d$ by renormalization-group method.
 
The SB t-J Hamiltonian is given by
\begin{eqnarray}
H&=&-t\sum_{x, i,\sigma} \left(b^{\dagger}_{x+ i}f^{\dagger}_{x 
\sigma}f_{x+ i \sigma}b_x+\mbox{H.c.}\right)\nonumber\\
&-&\frac{J}{2} \sum_{xi} \left| 
f^{\dagger}_{x\uparrow}f^{\dagger}_{x+ i\downarrow}
- f^{\dagger}_{x\downarrow}f^{\dagger}_{x+ i\uparrow}\right|^2.
\label{Hsb}
\end{eqnarray}
$f_{x\sigma}$ is the fermionic spinon operator with spin 
$\sigma (=\uparrow,\downarrow)$ at  
site $x$ of a 2d lattice \cite{2d}, and $b_{x}$ is the bosonic holon operator.
The direction  index  $ i ( = 1, 2)$ 
is used also as unit vectors.

The MFT is obtained by decoupling both $t$ and $J$ terms. For definiteness
we follow  Ubbens and Lee \cite{ul},
\begin{eqnarray}
H_{\rm MF}&=&\sum_{xi} 
\left\{ {3J\over 8}|\chi_{xi}|^2+{2\over 3J}|\lambda_{xi}|^2
\right\} \nonumber\\
&-&\sum_{xi}\left\{\chi_{xi}\left({3J\over 8}\sum_{\sigma} 
f^{\dagger}_{x+ i \sigma}
f_{x \sigma} 
+t b^{\dagger}_{x+ i}b_x\right) +\mbox{H.c.}\right\}  \nonumber \\
&-&{1\over 2}\sum_{x, i,\sigma} 
\left\{ \lambda_{xi} 
\left(f^{\dagger}_{x\uparrow}f^{\dagger}_{x+ i\downarrow}
- f^{\dagger}_{x\downarrow}f^{\dagger}_{x+ i\uparrow}\right)+\mbox{H.c.}
\right\}.
\label{Hdec1}
\end{eqnarray}
The partition function $Z(\beta) \; [\beta \equiv(k_B T)^{-1}]$  in
 path-integral formalism is given 
by integrating out two MF's, the hopping amplitude, $\chi_{xi}$, and 
RVB amplitude, $\lambda_{xi}$ 
defined on the link $(x, x+i)$ \cite{2body};

\begin{eqnarray}
Z &=& \int[db][df][d\chi][d\lambda]\exp A,\nonumber\\ 
A &\equiv& \int_0^{\beta}d\tau 
\left\{-\sum (b^{\dagger}_x\dot{b}_x +
\sum f^{\dagger}_{x\sigma}\dot{f}_{x\sigma}) - H_{\rm MF}\right\}.
\end{eqnarray}

In the CSS state ($T < T_{\rm CSS}$), 
$\langle \chi_{xi} \rangle \neq 0$.
The spin-gap state may realizes in CSS and is characterized by a 
condensation of $\lambda_{xi}$, 
$\langle \lambda_{xi} \rangle  \neq 0$.
Let us parameterize $ \chi_{xi} = \chi U_{xi}$, 
$U_{xi} \equiv \exp(i A_{xi})$, 
$\lambda_{xi} = (-)^i \lambda V_{xi}$, 
$V_{xi} \equiv \exp(i B_{xi})$, assuming uniform RVB.
If one ignores phase fluctuations by setting $A_{xi} = B_{xi} = 0$,
spinon excitations has the  energy $E(\mbox{\boldmath $k$})$,
$E^2(\mbox{\boldmath $k$}) 
=  \{( 3J\chi/4)   \sum_{i} \cos k_i-\mu_F\}^2 
+ \{ \lambda \sum_i (-)^i \cos k_i \}^2,$
where $\mu_F$ is the chemical potential to ensure $\langle   
f_{x\uparrow}^{\dagger}f_{x\uparrow} 
+ f_{x\downarrow}^{\dagger}f_{x\downarrow} \rangle = 1- \delta$
($\delta$ is  doping). We introduce also $\mu_B$ for
$\langle   b_{x}^{\dagger}b_{x} \rangle =  \delta$.
There appears a spin gap 
$  \lambda(\cos k_1  - \cos  k_2) $.

Under the gauge transformation, 
$b'_x = \exp(i\theta_x) b_x$, 
$f'_{x\sigma} = \exp(i\theta_x) f_{x\sigma}$,  
the phases of MF's transform as
$A'_{xi} = A_{xi} + \theta_x  -\theta_{x+ i}$, 
$B'_{xi} = B_{xi} + \theta_x +\theta_{x+ i}$. 
Their behavior can be studied by the effective lattice gauge 
theory $A_{\rm LGT}(U,V)$ that is obtained by integrating over 
 $b_x$ and $f_{x\sigma}$, e.g., by hopping expansion.
For $T^* < T < T_{\rm CSS}$, $A_{\rm LGT}$ contains only $U$'s 
interacting via the conventional gauge couplings like 
$\chi^4 U_{x\,2}\,U_{x+2\,1}\,U^{\dagger}_{x+1\,2}\,U^{\dagger}_{x\,1}$.
Their quadratic terms for $A_{xi}$ show  that $A_{xi}$ 
behaves as a massless 
gauge field. For $T_c < T < T^*$, $\lambda$ develops and new couplings like 
$\chi^2 \lambda^2 V^{\dagger}_{x+1\,2}\,U_{x+2\,1}V_{x\,2}\,U_{x\,1}$ 
are generated. This gives rise to a mass term of $A_{xi}$, 
$m^2_{A} A_{xi}^2$ with $m_{A}^2 = \chi^2 \lambda^2 $. 
Also, the quadratic $B_{xi}$ are also massive.
 
The holon part $A^{B}_{\rm eff}$ of the effective low-energy continuum 
field theory $A_{\rm eff}$ is obtained as 
\begin{eqnarray}
A_{\rm eff}^{B} & = & \int d\tau d^2x 
\left\{-\bar{b}\left(\partial_{\tau}-\mu_B\right)b 
-\frac{1}{2m_B}|D_i b|^2\right\}, 
\label{aeffb}
\end{eqnarray}
where $\ D_i \equiv \partial_i -i g A_i$ 
(we introduced the gauge coupling constant $g =1$ for convenience), 
and $(2m_B)^{-1} = \chi\;  t \; a^2$ ($a$ is the lattice spacing).
The spinon part $A^{F}_{\rm eff} $ may be written as
\begin{eqnarray}
A^{F}_{\rm eff}  
&\simeq& \int d\tau d^2x 
\left\{-\bar{f}\left(\partial_{\tau} -\mu_F\right)f 
- \frac{1}{2m_F} |D_i f|^2\right\} \nonumber\\
&&-\int d\tau d^2k  
\left\{\Delta_{\rm SG}(\mbox{\boldmath $k$}) 
\bar{f}_{\uparrow}(\mbox{\boldmath $k$},\tau)
\bar{f}_{\downarrow}(-\mbox{\boldmath $k$},\tau)
 +{\rm H.c.}\right\},
\label{aefff} 
\end{eqnarray}
with $(2m_F)^{-1} = 3J\chi a^2/8$.
Here we introduced the continuum version of the spin gap
\begin{equation}
  \Delta_{\rm SG}(\mbox{\boldmath $k$})
\equiv \pi(1-\delta)\hat{\lambda}(T)\frac{k_x^2 - k_y^2}{k_F^2},
\label{DelSG}
\end{equation}
where $k_F (\equiv\sqrt{2m_F\mu_F})$ is the Fermi momentum of spinons
and $k_F\simeq\sqrt{2\pi(1-\delta)}/a$
at $T\ll T_F$ ($T_F$ is the Fermi temperature).
In (\ref{DelSG}), we use the renormalized spin gap
$\hat{\lambda}(T)$ defined as   
$\langle  \lambda_{xi} \rangle = (-)^{i}\hat{\lambda}(T)$,
instead of its MF value $\lambda$ to take into account 
the effect of phase fluctuations of $\lambda_{xi}$ effectively.
Below we calculate the resistivity of the system 
$A_{\rm eff} = A^{B}_{\rm eff} + A^{F}_{\rm eff}$\cite{a0}.

The propagator of the gauge field,
$D_{ij}(\mbox{\boldmath $x$}, \tau)
\equiv \langle A_i(\mbox{\boldmath $x$}, \tau)A_j(0, 0)\rangle$, 
is generated by fluctuations of spinons and holons, i.e.,
$D_{ij}= (\Pi_{F}+\Pi_{B})^{-1}_{ij}$,
where 
\begin{eqnarray}
\Pi_{F,B\;ij}(\mbox{\boldmath $x$}, \tau), 
&\equiv& - \langle J_{F,B\;i}(\mbox{\boldmath $x$},\tau) 
J_{F, B\;j}(0,0)\rangle_{\rm 1PI}
\nonumber\\
& &+\delta_{ij}\delta(\mbox{\boldmath $x$})\delta(\tau)n_{F,B}, 
\end{eqnarray}
representing one-particle-irreducible (1PI)
diagrams of spinon and holon loops. 
$J_{{F}i} \equiv 
(2m_{F})^{-1}\sum_{\sigma}\{i\bar{f}_{\sigma}
\partial_i f_{\sigma}+{\rm H.c.}\}$ 
and $J_{Bi} \equiv (2m_{B})^{-1}\{i\bar{b}\partial_i b+{\rm H.c.}\}$
are currents  coupled to $A_{i}$, and 
$n_F = (1-\delta)/a^2$ and $n_B = \delta/a^2$.
In the Coulomb gauge,  the propagator at momentum 
$\mbox{\boldmath $q$}$ and  Matsubara frequency
$\epsilon_l \equiv 2\pi l /\beta$ is written as   
\begin{eqnarray}
D_{ij}(\mbox{\boldmath $q$}, \epsilon_l)&=& 
\left(\delta_{ij}-\frac{q_i q_j}{q^2}\right)
D(\mbox{\boldmath $q$}, \epsilon_l),
\nonumber\\
  D(\mbox{\boldmath $q$}, \epsilon_l) &=&
\left\{\Pi_{F}(\mbox{\boldmath $q$}, \epsilon_l)
+\Pi_{B}(\mbox{\boldmath $q$}, \epsilon_l)\right\}^{-1}.
\label{d}
\end{eqnarray}

Since we shall need $D$ later in  calculating $\rho(T)$,  we obtain $D$ below
in the random-phase approximation as $D \simeq (\Pi^R_B + \Pi^R_F)^{-1}$.
When the spin gap is sufficiently small, its effect to $\Pi^R_F$
 is evaluated by perturbation giving rise to  a mass term as discussed above; 
\begin{equation}
  \frac{\Pi^{R}_F(q, \epsilon_l)}{g^2}\simeq\left\{
    \begin{array}{rl}
      \frac{q^2}{12\pi m_F}
        +\sqrt{\frac{n_F}{2\pi}}\frac{|\epsilon_l|}{q}
        +\frac{n_F^S(T)}{m_F},
        & |\epsilon_l|\ll v_Fq\\
      \frac{n_F}{m_F},
      & |\epsilon_l| \gg v_Fq
    \end{array}
  \right. .
\label{pirf}  
\end{equation}
We used the relation, $\mu_F\simeq\pi n_F/m_F$
and $v_F\equiv k_F/m_F$.
The superfluid density of spinons 
is calculated  for small $\Delta_{\rm SG}(\mbox{\boldmath $k$})/(k_B T)$ as 
\begin{equation}
  n_F^S(T)\simeq \frac{n_F}{2\pi}\int d\phi 
\left|\frac{\Delta_{\rm SG} (\mbox{\boldmath $k$})}{2k_BT}\right|^2
= \frac{n_F}{2}\left|\frac{\pi(1-\delta)\hat{\lambda}(T)}{2k_BT}\right|^2,
\label{sfd}
\end{equation}
with $k_x /k_y = \tan \phi$ 
and $|\mbox{\boldmath $k$}| = k_F$.
$\Pi^R_B$ is given by
\begin{equation}
  \frac{\Pi^{R}_B(q, \epsilon_l)}{g^2}\simeq\left\{
    \begin{array}{rl}
      \frac{f_B(|\mu_B|)}{24\pi}\frac{q^2}{m_B}
        +\sqrt{\frac{n_B}{2\pi}}\frac{|\epsilon_l|}{q},
        & |\epsilon_l|\ll \frac{\sqrt{n_B}}{m_B}q\\
      \frac{n_B}{m_B},
      & |\epsilon_l| \gg \frac{\sqrt{n_B}}{m_B}q
    \end{array}
  \right.
\label{pirb}
\end{equation}
where $f_B(\epsilon)\equiv\left\{\exp(\beta \epsilon)-1\right\}^{-1}$.
Eqs.(\ref{pirf},\ref{pirb}) above are obtained for small $q (\ll \pi/a)$. 
For large $q$'s, they should be replaced by
anisotropic expressions due to  
$\Delta_{\rm SG}(\mbox{\boldmath $k$})$.  
These anisotropy can be ignored as long as the spin gap is sufficiently small.

From the linear-response theory and Ioffe-Larkin formula \cite{il},
the dc conductivity $\sigma (\equiv \sigma_{11} = \sigma_{22})$ 
is expressed as 
\begin{eqnarray}
&&\sigma_{ij}  
=  \lim_{\epsilon \rightarrow 0}\lim_{q\rightarrow 0}
\frac{e^2}{-i \epsilon}\tilde{\Pi}_{ij}(\mbox{\boldmath $q$},-i\epsilon), 
\nonumber\\  
&&\tilde{\Pi}_{ij}(\mbox{\boldmath $q$}, \epsilon) 
= \left\{\tilde{\Pi}_F^{-1}(\mbox{\boldmath $q$}, \epsilon) 
+ \tilde{\Pi}_B^{-1}(\mbox{\boldmath $q$}, \epsilon)\right\}^{-1}_{ij},  
\end{eqnarray}
where $\tilde{\Pi}$, $\tilde{\Pi}_{F,B}$ are response functions of
 electron,  spinon, and holon, respectively. 
So one has $ \sigma^{-1}  = \sigma_B^{-1}
+ \sigma_F^{-1}$. 
 
In the spin-gap state, the spinon conductivity 
diverges $\sigma_F \rightarrow \infty$ due to a superflow
generated by RVB condensation $\langle \lambda_{xi} \rangle \neq 0$.
This is an analog of the well-known fact in the BCS theory
that the electron resistivity vanishes below $T_c$ due to a
superflow generated by Cooper-pair condensation. Actually, 
$A_{\rm eff}^F$ has the same structure as the BCS model. 
Thus the total resistivity $\rho = \sigma^{-1}$ 
in the spin-gap state is equal to 
the resistivity of holons, $\rho = \sigma_B^{-1}$.
Effects of spinons to $\rho$ certainly exist and show up through
the dressed propagator $D(\mbox{\boldmath $q$}, \epsilon_l)$ 
in calculating $\tilde{\Pi}_B$. 

Now we calculate the response function $\tilde{\Pi}_B$. 
By solving the Schwinger-Dyson equation approximately following the  steps
in Ref.\cite{oim}, we arrive at 
\begin{eqnarray}
\tilde{\Pi}_{B\;ij}(0,\epsilon_l) &\simeq& \frac1\beta\sum_n
\int \frac{d^2q}{(2\pi)^2}\frac{q_iq_j}{{m_B}^2}R_B(q, \epsilon_n;\epsilon_l)
\nonumber\\
&&\times \ \ \frac{i\epsilon_l}{i\epsilon_l
-i\epsilon_l\Gamma_B(q, \epsilon_n; \epsilon_l)
-\Delta\Sigma_B(q, \epsilon_n; \epsilon_l)}, \nonumber\\
R_B(q, \epsilon_n;\epsilon_l)
&\equiv& G_B(q, \epsilon_n)G_B(q, \epsilon_n+\epsilon_l),
\nonumber\\
G_B(q, \epsilon_n)
&\equiv& \left(i\epsilon_n - \frac{q^2}{2m_B} + \mu_B\right)^{-1}.
\label{pib}
\end{eqnarray}

$\Delta\Sigma_B(q, \epsilon_n; \epsilon_l)$, representing 
diagrams containing self-energy of holons, $\Sigma_B(q, \epsilon_n)$, 
is necessary to keep gauge invariance,
\begin{eqnarray}
\Delta\Sigma_B(q, \epsilon_n; \epsilon_l)
&\equiv&  R_B^{-1}(q, \epsilon_n;\epsilon_l)
\{\Sigma_B(q, \epsilon_n)G_B^2(q, \epsilon_n) \nonumber\\
& &\quad -\Sigma_B(q, \epsilon_n+\epsilon_l)G_B^2(q, \epsilon_n+\epsilon_l)\},
\end{eqnarray}
However, in the perturbative calculation, this combination vanishes 
in the dc limit by the symmetry under summations. 
We expect this term dose not contribute to the dc resistivity,
and neglect it hereafter.
 
$\Gamma_B(q,\epsilon_n;\epsilon_l)$, representing vertex diagrams, 
contributes to $\sigma_B$,
\begin{eqnarray}
\Gamma_B(q,\epsilon_n;\epsilon_l)
&\equiv&\left(\frac{g}{m_B}\right)^2\frac1\beta\sum_{n'}
\int \frac{d^2q'}{(2\pi)^2} \nonumber\\
&&\times 
\left\{\frac{\mbox{\boldmath $q$}
\times (\mbox{\boldmath $q$}'-\mbox{\boldmath $q$})}
{|\mbox{\boldmath $q$}'-\mbox{\boldmath $q$}|}\right\}^2
\frac{\mbox{\boldmath $q$}\cdot(\mbox{\boldmath $q$}'
-\mbox{\boldmath $q$})}{q^2}
\nonumber\\
& &\times D(|\mbox{\boldmath $q$}'-\mbox{\boldmath $q$}|,
\epsilon_{n'}-\epsilon_n) R_B(q',\epsilon_{n'};\epsilon_l),
\label{gamma}
\end{eqnarray}
where $\mbox{\boldmath $q$}\times\mbox{\boldmath $q$}'\equiv q_xq'_y-q_yq'_x$.
We set $q$ of  $\Gamma_B$ in $\Pi_B$ to a fixed vector
of typical length $q_B$, $q_B^2 \equiv 8\pi n_B$\cite{qb}. 
We also fix the length
of $q'$ of $D$ in (\ref{gamma}) be $q_B$.
\begin{eqnarray}
\Gamma_B(q_B,\epsilon_n;\epsilon_l)
&\simeq&-\frac{g^2{q_B}^2}{8\pi^2m_B}\frac1\beta\sum_{n'}
    \int^{\pi}_{-\pi}d\phi\;\sin^2\phi
\nonumber\\  
& &\times D\left(q_B\sqrt{2(1-\cos\phi)},
 \epsilon_{n'}-\epsilon_n\right) \nonumber\\
& &\times \int^{\infty}_{|\mu_B|}dE\;\frac1{i\epsilon_{n'} -E}
\frac1{i\epsilon_{n'} +i\epsilon_l -E}.
\end{eqnarray}
We consider the underdoped region, 
$n_B \ll n_F \ (\delta \ll 1),$ 
and temperatures around $ \beta^{-1}\sim n_B/m_B$. 
Assuming that $D(q,\epsilon_l)$ in $\Gamma_B$ dominates in the region
near the static limit $\epsilon_l = 0$, we use the upper expressions
in (\ref{pirf},\ref{pirb}). In the denominator of $D$, the dissipation term, 
$\sqrt{\bar{n}/(2\pi)} |\epsilon_l|/q$, 
$\sqrt{\bar{n}} \equiv
 \sqrt{n_F} + \sqrt{n_B}$,
 is larger than the $q^2$ term, $q^2/(12\pi \bar{m})$,
$\bar{m}^{-1}  \equiv m_F^{-1} + f_B(|\mu_B|)/(2m_B)$,
as long as  $\epsilon_l \neq 0$, since their ratio is small,
$(q_B^2/\bar{m})/\{\sqrt{\bar{n} }/(q_B\beta)\} 
\sim {\cal O}(\sqrt{n_B/n_F})$. 

So the $n'$-sum is dominated at $\epsilon_{n'}=\epsilon_n$.
Then we get
\begin{eqnarray}
\Gamma_B(q_B,\epsilon_n;\epsilon_l)
&\simeq&-\frac{3 \bar{m}}{4\pi m_B}\frac1\beta
    \int^{\pi}_{-\pi}d\phi\frac{\sin^2\phi}
{(1-\cos\phi)+\frac{3 \bar{m}n_F^S(T)}{4m_Fn_B}}
\nonumber\\
& &\times
\left[\frac{\pi}{2\epsilon_l}
\left\{\mbox{sgn}(\epsilon_n+\epsilon_l)-\mbox{sgn}(\epsilon_n)\right\}
+{\cal O}(\epsilon_l^0)\right]\nonumber\\
&\simeq&
-\frac1{2\epsilon_l \tau(T)}
\left\{\mbox{sgn}(\epsilon_n+\epsilon_l)-\mbox{sgn}(\epsilon_n)\right\}
\label{gammab}
\end{eqnarray}
where
\begin{eqnarray}
\frac{1}{\tau(T)}
&\equiv& \frac{3\pi\bar{m}}{2m_B}\frac1\beta
 \Biggl[\left\{1+\frac{3\bar{m}n_F^S(T)}{4 m_F n_B}\right\}
\nonumber\\
& &\hspace{1.5cm}
-\sqrt{\left\{1+\frac{3\bar{m}n_F^S(T)}{4 m_F n_B}\right\}^2-1}
\Biggr].
\end{eqnarray}

To calculate $\tilde{\Pi}_{B\;ij}(0, \epsilon_l)$ we insert (\ref{gammab}) 
into (\ref{pib}) and do the $q$-integral and $n$-sum 
as in (\ref{gamma}) to get
\begin{eqnarray}
\tilde{\Pi}_{B\;ij}(0,\epsilon_l) &\simeq& \delta_{ij}\frac{n_B}{m_B}
\frac{i\epsilon_l}{ \tilde{C}(T)\epsilon_l +i\tau^{-1}(T)},
\end{eqnarray}
where $\lim_{\epsilon_l \rightarrow 0}\tilde{C}(T)$ is finite.

After analytic continuation $\epsilon_l > 0 \to -i\epsilon$
and using (\ref{sfd}),
we finally obtain the resistivity,
\begin{eqnarray}
\rho(T)& =& \frac{m_B}{e^2n_B}\frac{1}{ \tau(T)} \nonumber\\
&\propto&  
T \left\{ 1- \sqrt{\frac{3\bar{m}(1-\delta)^3}{4 m_F \delta}}
\left|\frac{\pi\hat{\lambda}(T)}{2k_BT}\right| \right\} 
+ {\cal O}( n_F^S ). 
\label{rho}
\end{eqnarray}
For $T^* < T < T_{\rm CSS}$, $n_F^S(T)= 0$ and
this result reproduces  the $T$-linear behavior of Ref.\cite{ln}.
For $T $ near and below $T^*$, 
one expects the  behavior
$|\hat{\lambda}(T)| \propto (T^*-T)^{d}$, with a critical exponent $d$,  
and we have 
$\rho(T) \propto T\{1-c(T^*-T)^{d}\}$. 
Note also that the downward deviation of $\rho(T)$ 
from the $T$-linear behavior is reduced 
with increasing the doping $\delta$.
In Fig.1, we  plot $\rho(T)$ of (\ref{rho}) with various values of $d$. 
The MFT value $d  = 1/2$ 
is not consistent with the experiment. To achieve smooth
deviation from the $T$-linear behavior, one needs $d > 1$.  
This suggests that fluctuation effect of phases of 
$\lambda_{xi}$ is important to obtain a 
realistic curve of $\rho(T)$.
One could produce a reliable curve of $\rho(T)$ by inserting experimental
data of $\hat{\lambda}(T)$ into (\ref{rho}), but the available 
experimental data are not enough for this purpose.  

The data \cite{uchida} show that one may fit $\rho$ in a form $C_0 + C_1 T$
for $  T^* < T$. This implies spinon contribution to $\rho$, 
calculated as  $  \sigma_F^{-1} 
\propto T^{4/3}/n_F$  \cite{ln}\cite{oim}, is negligibly small 
compared with  $\sigma_B^{-1} \propto T/n_B$ due to higher
power in $T$ and a small coefficient. 
 $\sigma_F^{-1} = 0 $ for $T < T^*$ as explained, 
but the discontinuity at $T=T^*$ in $\sigma_F^{-1}$ is not observable
due to its smallness.  

The constant part $C_0$, surviving below $T^*$,  may be attributed to 
scatterings of charged holons with impurities. 
They may be described by $H_{\rm imp} = \sum V_x b^{\dagger}_x b_x $,
where $V_x$ is a random potential. 
Actually, standard calculations show that it generates  
$T$-independent contribution to $\rho$,  
$\Delta \sigma_B^{-1} \propto 1/n_B  $ at
intermediate $T$'s \cite{c0}. 

As $T$ goes very low, $\Pi^{R}_F$ is dominated by the mass term,
$  \Pi^{R}_F(q, \epsilon_l)/g^2 \simeq  n_F^S(T)/m_F,$
while $\Pi^{R}_B$ does not change.
The Landau damping term from holons is smaller than the mass term above. 
This case has been studied in Ref.\cite{oim}, giving the result 
$\sigma_B^{-1} \propto T^2$ .
Thus, in $\rho(T)$,  weak-localization effect by impurities, $\rho_{\rm WL}
\sim C_{\rm WL} \log(T),$  dominates
over inelastic scatterings by gauge bosons.
This situation is in contrast with the effective gauge theory of 
two-dimensional electrons at half-filled Landau level,
in which the transverse mode of Chern-Simons gauge field remains massless
down to  $T = 0$ and renormalizes $C_{\rm WL}$ \cite{Khveshchenko}.


\begin{figure}
  \begin{picture}(210,210)
    \put(0,0){\epsfxsize 200pt \epsfbox{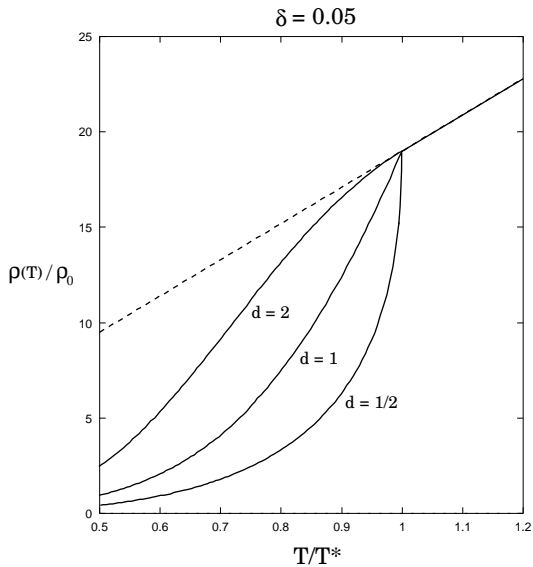}}
  \end{picture}
  \caption{Plot of the resistivity $\rho(T)$ divided by
  $\rho_0\equiv 2\pi k_BT^{*}m_F/(e^2n_F)$  for 
  $\hat{\lambda}(T)\simeq \lambda_0(1-T/T^{*})^d$ $(d = 1/2, 1, 2)$.
  For definiteness we chose $\delta=0.05$, $\lambda_0 = 2k_B T^{*}/ \pi$ 
  and $3\bar{m}/(4m_F) =0.5$.
  }
\end{figure}


\acknowledgments

M.Onoda thanks the Japan society for the promotion of science 
for financial support.



\begin{references}
%
\bibitem[*]{onoda} Electronic address: onoda@cms.phys.s.u-tokyo.ac.jp 
\bibitem[\dagger]{ichinose} Electronic address: ikuo@hep1.c.u-tokyo.ac.jp
\bibitem[\ddagger]{matsui} Electronic address: matsui@phys.kindai.ac.jp
%
%
\bibitem{sato}
See, e.g., T.Nishikawa {\it et al.},
J. Phys. Soc. Jpn. {\bf 63}, 1441 (1994) ; J.Takeda {\it et al.},
Physica {\bf C231}, 293 (1994); H.Y.Hwang {\it et al.}, Phys. Rev. Lett.
{\bf 72}, 2636 (1994). 
\bibitem{anderson}P.W.Anderson,
  Phys. Rev. Lett. {\bf 64}, 1839 (1990).
%
\bibitem{imcss}
I.Ichinose and T.Matsui, 
Nucl. Phys. {\bf B394}, 281 (1993);  
Phys. Rev. {\bf B51}, 11860 (1995).
N. Nagaosa, Phys. Rev. Lett. {\bf 71} (1993) 4210, also studied
a CD transition of certain dissipative gauge theory of fermions. 
%
\bibitem{fiory}
M.Gurvitch and A.T.Fiory, Phys. Rev. Lett. {\bf 59}, 1337 (1987).
%
\bibitem{ln}P.A.Lee and N.Nagaosa, Phys. Rev. {\bf B46}, 5621 (1992).
%
\bibitem{uchida}
T.Ito, K.Takenaka and S.Uchida, Phys. Rev. Lett. {\bf 70}, 3995 (1993). 
\bibitem{sgexp}
H.Yasuoka et al.,"Strong Correlation and Superconductivity", ed. 
H.Fukuyama et al. (Springer Series, Berlin, 1989) p.254;
M.Takigawa et al., Phys. Rev. {\bf B43}, 247 (1991);
J.Rossat-Mignot et al., Physica {\bf 185-189C}, 86 (1991);
J.M.Tranquada,et al., Phys. Rev. {\bf B46}, 5561 (1992).
%
\bibitem{2d}
Although we do explicit calculations in two dimensions, 
we keep in mind that  weak but finite three
dimensionality is necessary to stabilize the spin gap.
%
\bibitem{ul}
M.U.Ubbens and P.A.Lee, Phys. Rev. {\bf B46}, 8434 (1992); 
{\it ibid.} {\bf B49}, 6853 (1994).
%
\bibitem{2body}
To reproduce the exact $Z$, one must add extra two-body terms
in the action, whose effects are neglected below as in usual MFT.
They reduce $T_{\rm CSS}$ as $\delta$ increases.
Also we shall ignore the $\tau$-dependence of $\chi_{xi}$ and $\lambda_{xi}$.
%
\bibitem{a0} 
The SB local constraint can be  incorporated by inserting
the time-component $A_0$ of gauge field in (\ref{aeffb}, \ref{aefff}).
However, its fluctuation effects are negligible at 
$T < T_{\rm CSS}$
since they are short-ranged.
%
\bibitem{il}
L.B.Ioffe and I.Larkin, Phys.Rev. {\bf B39},8988 (1989).
%
\bibitem{oim} 
M. Onoda, I.Ichinose, and T.Matsui, Phys. Rev. {\bf B54}, 13674 (1996). 
%
\bibitem{qb}
This $q_B$ is determined so that this procedure applied for a 
similar  integral, (\ref{gamma}) with $D =1$,
gives rise to the exact result.
%
\bibitem{c0}
See also P.A.Lee and N.Nagaosa, Phys. Rev. Lett. {\bf 79}, 3755 (1997).
\bibitem{Khveshchenko}
D.V.Khveshchenko, Phys. Rev. Lett. {\bf  77}, 362 (1996).

\end{references}
\end{document}